\documentclass[preprint,showpacs,preprintnumbers,amsmath,amssymb]{revtex4}

\usepackage{graphicx}
\usepackage{dcolumn}
\usepackage{bm}

\font\scripti=cmmi7
\font\scriptscripti=cmmi5
\def\sib#1{\setbox0 = \hbox{\scripti #1}
  \kern-.02em\copy0\kern-\wd0
  \kern.04em\box0} 
\def\ssib#1{\setbox0 = \hbox{\scriptscripti #1}
  \kern-.02em\copy0\kern-\wd0
  \kern.04em\box0} 
\font\tenib=cmmib10 
\skewchar\tenib='177 \skewchar\tenib='177 \skewchar\tenib='177
\def\pbold#1{\setbox0 = \hbox{$ #1 $}
  \kern-.022em\copy0\kern-\wd0
  \kern.011em\copy0\kern-\wd0
  \kern.011em\copy0\kern-\wd0
  \kern.011em\copy0\kern-\wd0
  \kern.011em\box0} 
  
\def\lesssim{\ \raise.3ex\hbox{$<$}\kern-0.8em\lower.7ex\hbox{$\sim$}\ }
\def\gesim{\ \raise.3ex\hbox{$>$}\kern-0.8em\lower.7ex\hbox{$\sim$}\ }

\begin{document}
\title{Proposed method to realize the $p$-wave superfluid state using a $s$-wave superfluid Fermi gas with a synthetic spin-orbit interaction}
\author{T. Yamaguchi}\author{Y. Ohashi}
\affiliation{Department of Physics, Keio University, 3-14-1 Hiyoshi, Yokohama 223-8522, Japan}
\date{\today}

\begin{abstract}
We theoretically propose an idea to reach the $p$-wave superfluid phase in an ultracold Fermi gas. The key of our idea is that the pairing symmetry 
of a Fermi superfluid is fully dominated by the symmetry of the superfluid order parameter, which is essentially given by the product of a pair 
amplitude and a pairing interaction. Noting this, in our proposal, we first prepare a $p$-wave pair amplitude by, not using a $p$-wave interaction, 
but using the phenomenon that a $p$-wave pair amplitude is induced in an $s$-wave superfluid Fermi gas with antisymmetric spin-orbit interaction. 
In this case, although the system is still in the $s$-wave superfluid state with the $s$-wave superfluid order parameter, when one suddenly replaces 
the $s$-wave interaction by an appropriate $p$-wave one (which is possible in cold Fermi gases by using a Feshbach resonance technique), the product 
of the $p$-wave interaction and the $p$-wave pair amplitude that has already been prepared in the spin-orbit coupled $s$-wave superfluid state 
immediately gives a finite $p$-wave superfluid order parameter. Thus, at least just after this manipulation, the system is in the $p$-wave superfluid 
state, being characterized by the artificially produced $p$-wave superfluid parameter. In this paper, to assess our idea, we evaluate the $p$-wave pair 
amplitude in a spin-orbit coupled $s$-wave superfluid Fermi gas at $T=0$. We determine the region where a large $p$-wave pair amplitude is obtained in 
the phase diagram with respect to the strengths of the $s$-wave pairing interaction and the spin-orbit coupling. We also discuss the accessibility of 
this optimal region on the viewpoint of the superfluid phase transition temperature. Since the achievement of a $p$-wave superfluid Fermi gas is one of 
the most crucial issues in cold atom physics, our proposal would be useful for this exciting challenge.
\end{abstract}

\maketitle
\par
\section{INTRODUCTION}
\par
Since the achievement of the $s$-wave superfluid phase transition in $^{40}$K\cite{Regal} and $^6$Li\cite{Zwierlein1,Kinast,Bartenstein} Fermi gases, 
the possibility of the $p$-wave superfluid state has extensively been discussed as the next challenge in cold Fermi gas 
physics\cite{OhashiP,Ho,Gurarie,Gurarie2,Levinsen,Botelho,Iskin,Iskin2,Iskin3,Grosfeld,Mizushima,Han,Cheng,Maier,Inotani}. Although no one has succeeded 
in this attempt, $p$-wave Feshbach resonances have already been discovered in $^{40}$K\cite{Regal2a,Ticknor,Gunter,Gaebler1} and $^6$Li\cite{Zhang,Schunck,Mukaiyama}. 
Thus, once a $p$-wave superfluid Fermi gas is obtained, as in the $s$-wave case\cite{Eagles,Leggett,Leggett2,NSR,SadeMelo,Ohashi,Strinati,Levin,Giorginia,Bloch,Zwerger}, 
one can systematically study this unconventional Fermi superfluid from the weak-coupling regime to the strong-coupling limit, by adjusting the threshold energy of a 
$p$-wave Feshbach resonance\cite{Timmermans,Chin}. Since this pairing state has also been discussed in various fields, such as liquid $^3$He\cite{pwave-order,Vollhardt,Mineev}, 
metallic superconductivity\cite{Sigrist,UPt3,SrRuO4}, as well as neutron star\cite{neutron}, the realization of a tunable $p$-wave superfluid Fermi gas would also give a big 
impact in these research fields.
\par
In the current stage of research toward the realization of a $p$-wave superfluid Fermi gas, one serious difficulty is that, although a $p$-wave attractive interaction is 
necessary for this pairing state, it also causes the so-called three-body loss\cite{Levinsen,3loss2,3loss3}, as well as the dipolar relaxation\cite{Gaebler1}, leading to very 
short lifetime of $p$-wave pairs\cite{Chevy}. Thus, although one can prepare $p$-wave molecules\cite{Gaebler1,Regal2,Inaba,Fuchs}, they are soon destroyed, before the $p$-wave 
condensate grows. Thus, at this stage, it is crucial to overcome this difficulty.
\par
The purpose of this paper is to propose a possible route to reach the $p$-wave superfluid phase in an ultracold Fermi gas. To explain our idea in a simple manner, we first note 
that any Fermi superfluid is characterized by a superfluid order parameter $\Delta_{\sigma\sigma'}({\bm p})$, being consisting of the product of the pairing interaction 
$U({\bm p},{\bm p}')$ and the pair amplitude $\Phi_{\sigma,\sigma'}({\bm p})=\langle c_{{\bm p},\sigma}c_{-{\bm p},\sigma'}\rangle$ as
\begin{eqnarray}
\Delta_{\sigma,\sigma'}({\bm p})
&=&\sum_{\bm p'}U({\bm p},{\bm p'})
\langle c_{{\bm p'},\sigma}c_{-{\bm p'},\sigma'}\rangle
\nonumber
\\
&=&
{\bar U}\gamma_{\bm p}\sum_{\bm p'}\gamma_{\bm p'}
\Phi_{\sigma,\sigma'}({\bm p}').
\label{eq.1}
\end{eqnarray}
Here, $c_{{\bm p},\sigma}$ is the annihilation operator of a fermion with momentum ${\bm p}$ and (pseudo) spin $\sigma=\uparrow,\downarrow$. In the second line in Eq. (\ref{eq.1}), we 
have assumed the separable interaction $U({\bm p},{\bm p'})={\bar U}\gamma_{\bm p}\gamma_{{\bm p}'}$ (where ${\bar U}$ is a coupling constant and $\gamma_{\bm p}$ is a basis function). 
When the pair amplitude $\Phi_{\sigma,\sigma'}({\bm p})$ has the same symmetry as that of the basis function $\gamma_{\bm p}$, we obtain a finite value of the superfluid order parameter 
$\Delta_{\sigma,\sigma'}({\bm p})\propto \gamma_{\bm p}$. 
\par
In a Fermi superfluid, the pair amplitude $\Phi_{\sigma,\sigma'}({\bm p})$ is usually produced by the pairing interaction $U({\bm p},{\bm p}')$ of the system. Thus, current experiments 
usually deal with a Fermi gas with a $p$-wave pairing interaction from the beginning. Although this approach seems the shortest way to reach the $p$-wave superfluid state, at present, 
no one has succeeded in this attempt, because of the above mentioned serious problem. 
\par
Instead of this ordinary approach, we consider an alternative route. That is, we first prepare only a $p$-wave pair amplitude in a system with {\it no} $p$-wave interaction. Of course, 
even when we prepare this quantity, the system is still not a $p$-wave Fermi superfluid, because the $p$-wave superfluid order parameter is absent due to the vanishing $p$-wave interaction. 
However, during this preparation, various difficulties originating from the $p$-wave interaction can be avoided. After this preparation, when the $s$-wave pairing interaction is suddenly 
replaced by a $p$-wave one $U_{p}({\bm p},{\bm p}')={\bar U}_{p}\gamma_{\bm p}\gamma_{{\bm p}'}$ (where the $p$-wave basis function $\gamma_{\bm p}$ is chosen so that the momentum summation 
in Eq. (\ref{eq.1}) can be finite), the product of this interaction and the $p$-wave pair amplitude that has been prepared in advance immediately gives a finite value of the $p$-wave superfluid 
order parameter $\Delta_{\sigma,\sigma'}({\bm p})\propto\gamma_{\bm p}$. Thus, at least just after this manipulation, the system is regarded as a $p$-wave Fermi superfluid, by definition. 
\par
In a sense, the superfluid order parameter is artificially produced in our approach, so that the resulting $p$-wave superfluid Fermi gas would initially be in the non-equilibrium state. 
In addition, when the $p$-wave interaction is turned on, the above mentioned problem occurs, so that the $p$-wave superfluid phase might eventually be destroyed. However, the $p$-wave 
superfluid order parameter would continue to exist for a while, during the decay and relaxation of the system. This is quite different from the ordinary approach, where the system is destroyed 
before the reach of the $p$-wave superfluid phase. 
\par
To prepare a $p$-wave pair amplitude in our approach, the recent artificial gauge field would be useful\cite{SOC0,Rb-SOC1,Rb-SOC2,Rb-SOC3,Rb-SOC4,Li-SOC,K-SOC,coldSOC5,coldSOC6}. 
This sophisticated technique enables us to introduce an antisymmetric spin-orbit interaction to an ultracold atom gas, leading to the broken spatial inversion symmetry. In this case, 
the parity is no longer a good quantity to classify the spatial pairing symmetry of the Fermi superfluid order parameter\cite{Fujimoto,noncen}. Since the pair wavefunction must be antisymmetric 
with respect to the exchange of two fermions, the parity mixing naturally leads to the admixture of the spin singlet and spin triplet states\cite{Fujimoto,noncen}. As a result, even in a 
$s$-wave Fermi superfluid, the pair amplitude may have the spin-triplet component. Thus, using this, one may prepare a $p$-wave pair amplitude without relying on a $p$-wave interaction. 
Indeed, Refs.\cite{Hu2011,2B1} explicitly show that the $p$-wave pair amplitude is induced in a $s$-wave superfluid Fermi gas in the presence of the Rashba spin-orbit interaction\cite{Rashba}, 
described by the Hamiltonian,
\begin{equation}
H_{\rm Rashba}=\lambda_{\rm Rashba} \sum_{\bm p}[p_x{\hat \sigma}_y^{\alpha,\alpha'}-p_y{\hat \sigma}_x^{\alpha,\alpha'}]c_{{\bm p},\alpha}^\dagger c_{{\bm p},\alpha'},
\label{eq.3}
\end{equation}
where ${\hat \sigma}_i (i=x,y,z)$ are Pauli matrices acting on spin space and $\lambda_{\rm Rashba}$ is a spin-orbit coupling constant. At present, a simple antisymmetric spin-orbit interaction 
has been realized in $^6$Li\cite{Li-SOC} and $^{40}$K\cite{K-SOC} Fermi gases and various ideas to synthesize more complicated spin-orbit couplings have also been proposed\cite{coldSOC5,coldSOC6}.
\par
Once a $p$-wave pair amplitude is induced in a $s$-wave superfluid Fermi gas, the $p$-wave superfluid order parameter can be produced by the sudden replacement of the $s$-wave interaction with a 
$p$-wave one, by tuning an external magnetic field from a $s$-wave Feshbach resonance field to a $p$-wave Feshbach resonance field. After this manipulation, the $s$-wave superfluid order parameter immediately 
vanishes, because of the vanishing $s$-wave interaction. Although the $s$-wave pair amplitude still remains, since the pairing symmetry of a Fermi superfluid is determined by the symmetry of the order 
parameter, this system is in the $p$-wave superfluid state.
\par
We note that the admixture of the spin-singlet pairing and spin-triplet pairing by an antisymmetric spin-orbit interaction has also been discussed in non-centrosymmetric superconductivity\cite{noncen}. 
For example, Ref.\cite{noncen2} points out the importance of this admixture to understand the anomalous behavior of the penetration depth observed in the non-centrosymmetric superconductor Li$_2$PtB$_3$.
\par
Our idea is somehow related to the proximity effect in a superconductor-normal metal (S-N) junction\cite{deGennes}. In this case, the pair amplitude in the S side penetrates into the N side, so that, 
when it couples with an interaction existing in the N side, a finite superconducting order parameter appears in the N side. Since the pair amplitude in the N side is fully supplied from the S side, 
even when the interaction in the N side is {\it repulsive}, this proximity-induced superconducting state is obtained\cite{Ohashi1996}.
\par
In this paper, to assess our idea, we investigate a $s$-wave superfluid Fermi gas with an antisymmetric spin-orbit interaction. Using the strong-coupling theory developed by Eagles\cite{Eagles} 
and Leggett\cite{Leggett,Leggett2} (which is sometimes referred to the BCS-Leggett theory in the literature), we examine how large a $p$-wave pair amplitude is induced by a spin-orbit interaction 
at $T=0$. We determine the region where a large $p$-wave amplitude is obtained, in the phase diagram with respect to the strengths of the $s$-wave interaction and the spin-orbit coupling. We also 
examine the accessibility of this ``optimal region" within the current experimental technique. For this purpose, employing the strong-coupling theory developed by Nozi\`eres and Schmitt-Rink (NSR)\cite{NSR}, 
we calculate the superfluid phase transition temperature $T_{\rm c}$ around this region. We briefly note that $T_{\rm c}$ in a spin-orbit coupled ultracold Fermi gas has recently been discussed 
in the BCS (Bardeen-Cooper-Schrieffer)-BEC (Bose-Einstein condensation) region within the framework of a $T$-matrix approximation\cite{Hu2,Zheng}, where pairing fluctuations are treated within 
the static approximation\cite{Levin}. In this paper, we do not employ the static approximation, but fully take into account dynamical properties of pairing fluctuations within the NSR scheme. 
\par
For the time evolution of the system after the introduction of a $p$-wave interaction, we need to deal with a non-equilibrium Fermi gas, which we will separately discuss in our future paper.
\par
This paper is organized as follows. In Sec. II, we explain the outline of our formulations at $T=0$, as well as at $T_{\rm c}$, in the BCS-BEC crossover region. Here, we also introduce the condensate 
fraction, as a useful quantity to evaluate the magnitude of the pair amplitude. In Sec. III, we discuss how large a $p$-wave pair amplitude is induced at $T=0$, when the inversion symmetry of the system 
is broken by a spin-orbit interaction. We clarify the condition to obtain a large $p$-wave pair amplitude. We also calculate the superfluid phase transition temperature $T_{\rm c}$, 
to examine the accessibility of this condition within the current experimental technique. Throughout this paper, we set $\hbar=k_{\rm B}=1$, and the system volume is taken to be unity, for simplicity.

\par
\par
\section{Formulation}
\par
We consider a spin-orbit coupled uniform two-component Fermi gas with an $s$-wave pairing interaction. To explain our formulation at $T=0$ within the framework of the BCS-Leggett 
theory\cite{Eagles,Leggett,Leggett2}, as well as the formulation at $T_{\rm c}$ within the framework of the NSR theory\cite{NSR}, in a unified manner, the functional integral formalism 
is convenient\cite{SadeMelo,Negele}. The partition function in this formalism is given by
\begin{equation}
\mathcal{Z}=\int\prod_{\sigma}\mathcal{D}{\bar \Psi}_\sigma\mathcal{D}\Psi_\sigma
e^{-S},
\label{eq.4}
\end{equation}    
where the action $S$ has the form,     
\begin{equation}
S= \int dx
\Bigl[
\sum_\sigma
{\bar \Psi}_\sigma(x)
\Bigl[
{\partial \over \partial\tau}
+
{{\hat {\bm p}}^2 \over 2m}-\mu
\Bigr]
\Psi_{\sigma}(x)
+
\sum_{\sigma,\sigma'}
{\bar \Psi}_\sigma(x)
h_{\rm so}^{\sigma,\sigma'}
\Psi_{\sigma'}(x)
-U_{s}
{\bar \Psi}_\uparrow(x){\bar \Psi}_\downarrow(x)
\Psi_\downarrow(x)\Psi_\uparrow(x)
\Bigr].
\label{eq.5} 
\end{equation}
Here, $x=({\bm r},\tau)$ and $\int dx=\int_0^\beta d\tau\int d{\bm r}$, where $\beta=1/T$. $\Psi_\sigma({\bm r},\tau)$ and ${\bar \Psi}_\sigma({\bm r},\tau)$ are a Grassmann variable and its conjugate, 
describing Fermi atoms with the atomic mass $m$ and pseudospin $\sigma=\uparrow,\downarrow$, specifying two atomic hyperfine states. ${\hat {\bm p}}=-i\nabla$ is the momentum operator in real space 
and $\mu$ is the Fermi chemical potential. The $s$-wave pairing interaction $-U_{s}~(<0)$ is assumed to be tunable by an $s$-wave Feshbach resonance. As usual, we measure the interaction 
strength in terms of the $s$-wave scattering length $a_{s}$, given by
\begin{equation}
{4\pi a_{s} \over m}
=-{U_{s} \over 1-U_{s}\sum_{\bm p}^{p_{\rm c}}
{1 \over 2\varepsilon_{\bm p}}},
\label{eq.1b}
\end{equation}
where $\varepsilon_{\bm p}=p^2/(2m)$ and $p_{\rm c}$ is a cutoff momentum.
\par
The antisymmetric spin-orbit interaction ${\hat h}_{\rm so}=\{h_{\rm so}^{\sigma,\sigma'}\}$ with linear-momentum dependence in Eq. (\ref{eq.5}) generally has the form\cite{rashbon1,rashbon2,2B2,2B3}
\begin{equation}
{\hat h}_{\rm so}=\sum_{i,j}{\hat p}_i\lambda_{i,j}{\hat \sigma}_j,
\label{eq.6}
\end{equation}
where the $3\times 3$-matrix ${\hat \lambda}=\{\lambda_{i,j}\}$ ($i,j=x,y,z$) describes spin-orbit coupling strengths. However, this paper deals with the simpler version, 
${\hat \lambda}={\rm diag}[\lambda_\perp,\lambda_\perp,\lambda_z]$, that is,
\begin{equation}
{\hat h}_{\rm so}=
\lambda_{\bot}[\hat{p}_x{\hat \sigma}_x+\hat{p}_y{\hat \sigma}_y]
+\lambda_z\hat{p}_z{\hat \sigma}_z.
\label{eq.7}
\end{equation}
Here, we take $\lambda_{\bot}, \lambda_{z}\ge 0$ without loss of generality. Although Eq. (\ref{eq.7}) cannot describe all possible spin-orbit interactions, it still covers some typical cases 
that have recently been discussed in cold atom physics. The Rashba-type spin-orbit interaction in Eq. (\ref{eq.3}) is obtained by setting $\lambda_z=0$ and rotating the momentum space by $\pi/2$ 
around the $p_z$ axis. The single-component spin-orbit interaction $h_{\rm so}\sim {\hat p}_x{\hat \sigma}_y$, which has recently been synthesized in $^6$Li\cite{Li-SOC} and $^{40}$K\cite{K-SOC} 
Fermi gases, is also obtained by setting $\lambda_z=0$ and rotating the momentum space by $\pi/2$ around the $p_y$ axis, which is followed by the rotation in the $({\hat \sigma}_x,{\hat \sigma}_y,{\hat \sigma}_z)$ 
space by $\pi/2$ around the ${\hat \sigma}_x$ axis. 
\par
As usual, we introduce the Cooper pair field $\Delta(x)$ and its conjugate $\Delta^*(x)$, using the Hubbard-Stratonovich transformation\cite{SadeMelo}. Carrying out the functional integrals with 
respect to $\Psi_\sigma(x)$ and ${\bar \Psi}_\sigma(x)$, one has\cite{SadeMelo}
\begin{equation}
\mathcal{Z}=\int \mathcal{D}\Delta^*\mathcal{D}\Delta e^{-S_{\rm eff}(\Delta,\Delta^*)},
\label{eq.8}
\end{equation}  
where the action $S_{\rm eff}(\Delta,\Delta^*)$ is given by
\begin{equation}
S_{\rm eff}(\Delta,\Delta^*)= \int dx
{|\Delta(x)|^2 \over U_{s}}-{1 \over 2}\rm{Tr}\ln[-{\hat G}^{-1}].
\label{eq.9}
\end{equation}
Here,
\begin{equation}
{\hat G}^{-1}(x,x')=
\left(
\begin{array}{cc}
\displaystyle
-{\partial \over \partial\tau}-
\Bigl[{{\hat {\bm p}}^2 \over 2m}-\mu\Bigr]-{\hat h}_{\rm so} & 
\displaystyle
i{\hat \sigma}_y\Delta(x) \\
\displaystyle
-i{\hat \sigma}_y\Delta^*(x)  
& 
\displaystyle
-{\partial \over \partial\tau}+
\Bigl[{{\hat {\bm p}}^2 \over 2m}-\mu\Bigr]
-{\hat h}_{\rm so}^*
\end{array}
\right)
\delta(x-x')
\label{eq.10} 
\end{equation}
is the inverse of the $4\times 4$-matrix single-particle thermal Green's function ${\hat G}(x,x^{\prime})=-\langle {\rm T}_{\tau}\{{\hat \Psi}(x){\hat \Psi}^\dagger(x')\}\rangle$ in the operator formalism, where 
\begin{eqnarray}
{\hat \Psi}(x)=
\left(
\begin{array}{c}
\Psi_{\uparrow}(x) \\
\Psi_{\downarrow}(x) \\
\Psi_{\uparrow}^{\dagger}(x) \\
\Psi_{\downarrow}^{\dagger}(x)
\end{array}
\right)
\label{eq.10b}
\end{eqnarray}
is the four component Nambu field\cite{Maki,Shiba}. 
\par
\par
\subsection{$p$-wave pair amplitude in a spin-orbit coupled $s$-wave superfluid state at $T=0$}
\par
We evaluate the $p$-wave pair amplitude at $T=0$ within the framework of the BCS-Leggett\cite{Eagles,Leggett,Leggett2} theory. In the functional integral formalism, this strong-coupling theory is simply 
reproduced by approximately evaluating the functional integral in the partition function $\mathcal{Z}$ in Eq. (\ref{eq.8}) by the value at the saddle point solution ($\Delta_{s}$), which is determined 
from the equation,
\begin{equation}
0=
\Bigl(
{\delta S_{\rm eff} \over \delta\Delta^*(x)}
\Bigr)_{\Delta(x)=\Delta^*(x)=\Delta_{s}}
=
{\Delta_{s} \over U_{s}}
+{1 \over 4\beta}
\sum_{{\bm p},\omega_n}
{\rm Tr}
\Bigl[[{\hat \rho}_y+i{\hat \rho}_x]
{\hat \sigma}_y {\hat G}^{\rm MF}({\bm p},i\omega_n)\Bigr],
\label{eq.11}
\end{equation}
where $\omega_n$ is the fermion Matsubara frequency and ${\hat \rho}_j$ ($j=x,y,z$) are Pauli matrices acting on particle-hole space. (Note that ${\hat \sigma}_j$ ($j=x,y,z$) act on spin space.) 
In Eq. (\ref{eq.11}), ${\hat G}^{\rm MF}({\bm p},i\omega_n)$ is the Fourier transformed Green's function in the mean-field BCS approximation, given by
\begin{eqnarray}
{\hat G}^{\rm MF}({\bm p},i\omega_n)
&=&
{1 \over
i\omega_n-[\xi_{\bm p}+{\bm p}_\lambda\cdot{\hat {\bm \tau}}]{\hat \rho}_z
-{\hat \rho}_y{\hat \sigma}_y\Delta_{s}
}
\nonumber
\\
&=&
-{1 \over 2}\sum_{\alpha=\pm}
{i\omega_n+[\xi_{\bm p}+{\bm p}_\lambda\cdot{\hat {\bm \tau}}]\rho_z+\Delta_{s}
{\hat \rho}_y{\hat \sigma}_y
\over \omega_n^2+(E_{\bm p}^\alpha)^2
}
\Bigl[
\bf{1}+\alpha{{\bm p}_\lambda\cdot{\hat {\bm \tau}} \over |{\bm p}_\lambda|}
\Bigr], 
\label{eq.11b}
\end{eqnarray}
where $\xi_{\bm p}=\varepsilon_{\bm p}-\mu=p^2/(2m)-\mu$ is the kinetic energy of a Fermi atom, measured from the Fermi chemical potential $\mu$. 
$E_{\bm p}^\alpha=\sqrt{(\xi_{\bm p}^\alpha)^2+\Delta_{s}^2}$ describes the Bogoliubov single particle excitations, where 
$\xi_{\bm p}^\alpha=\xi_{\bm p}+\alpha|{\bm p}_\lambda|$ with ${\bm p}_\lambda=(\lambda_\perp p_x,\lambda_\perp p_y,\lambda_zp_z)$. In Eq. (\ref{eq.11b}), 
${\hat {\bm \tau}}=({\hat \rho_z}{\hat \sigma}_x,{\hat \sigma}_y,{\hat \rho}_z{\hat \sigma}_z)$ is the spin operator under the four component Nambu representation\cite{Maki,Shiba}. 
Summing up the Matsubara frequencies in Eq. (\ref{eq.11}), we obtain the BCS gap equation in the presence of spin-orbit interaction. At $T=0$, it has the form,
\begin{eqnarray}
1={U_{s} \over 2}\sum_{{\bm p},\alpha=\pm}{1 \over 2E_{\bm p}^\alpha}.
\label{eq.12}
\end{eqnarray}
Eliminating the ultraviolet divergence from Eq. (\ref{eq.12}), we obtain\cite{SadeMelo,Leggett} 
\begin{equation}
1=-{4\pi a_{s} \over m}
\sum_{\bm p}
\Bigl[{1 \over 2}\sum_{\alpha=\pm}
{1 \over 2E_{\bm p}^\alpha}-{1 \over 2\varepsilon_{\bm p}}
\Bigr].
\label{eq.14}
\end{equation}
\par
We solve the renormalized gap equation (\ref{eq.14}), together with the equation for the number $N$ of Fermi atoms, which is obtained from the mean-field thermodynamic potential 
$\Omega_{\rm MF}=TS_{\rm eff}(\Delta_{s},\Delta_{s})$ as
\begin{equation}
N = -{\partial\Omega_{\rm MF} \over \partial\mu}
=
{1 \over 2\beta}\sum_{{\bm p},\omega_n}
{\rm Tr}
\Bigl[
\rho_z{\hat G}^{\rm MF}({\bm p},i\omega_{n})e^{i\rho_z\omega_n\delta}
\Bigr]
=\sum_{\bm p}
\left[
1-{1 \over 2}\sum_{\alpha=\pm}
{\xi_{\bm p}^{\alpha} \over E_{\bm p}^\alpha}
\right]
\label{eq.15}
\end{equation}
(where $\delta$ is an infinitesimally small positive number), to self-consistently determine $\Delta_{s}$ and $\mu$.
\par
The pair amplitude $\Phi({\bm p},S,S_z)$ with total spin $S$ and its $z$-component $S_z$ is obtained from the off-diagonal components of the Green's function ${\hat G}^{\rm MF}({\bm p},i\omega_n)$ 
in Eq. (\ref{eq.11b}). Noting that ${\hat G}^{\rm MF}({\bm p},\tau)$ in the operator formalism has the form
\begin{eqnarray}
{\hat G}^{\rm MF}({\bm p},\tau)=-
\langle
{\rm T}_\tau
\{
\left(
\begin{array}{c}
c_{{\bm p},\uparrow}(\tau) \\
c_{{\bm p},\downarrow}(\tau) \\
c^\dagger_{-{\bm p},\uparrow}(\tau) \\
c^\dagger_{-{\bm p},\downarrow}(\tau)
\end{array}
\right)
(
c^\dagger_{{\bm p},\uparrow}(0), 
c^\dagger_{{\bm p},\downarrow}(0),
c_{-{\bm p},\uparrow}(0),
c_{-{\bm p},\downarrow}(0)
)\}
\rangle,
\label{eq.16}
\end{eqnarray}
one finds,
\begin{eqnarray}
\Phi({\bm p},0,0)
&=&
{1 \over 2}
\Bigl[
\langle 
c_{{\bm p},\uparrow} c_{-{\bm p},\downarrow}
\rangle
-
\langle
c_{{\bm p},\downarrow} c_{-{\bm p},\uparrow}
\rangle
\Bigr]
=-
{1 \over 2}
\Bigl[
{\hat G}^{\rm MF}_{14}({\bm p},\tau=0)
-
{\hat G}^{\rm MF}_{23}({\bm p},\tau=0)
\Bigr]
\nonumber\\
&{ }&~~~~~~~~~~~~~~~~~~~~~~~~~~~~~~~~~~~~~~
=-{1 \over 2}
\sum_{\alpha=\pm}{\Delta_{s} \over 2E_{\bm p}^\alpha},
\nonumber
\\
\Phi({\bm p},1,1)
&=&
\langle 
c_{{\bm p},\uparrow} c_{-{\bm p},\uparrow}
\rangle
=-{\hat G}^{\rm MF}_{13}({\bm p},\tau=0)
={\lambda_\perp \over 2}
{p_x-ip_y \over |{\bm p}_\lambda|}
\sum_{\alpha=\pm}{\alpha\Delta_{s} \over 2E_{\bm p}^\alpha}, 
\nonumber
\\
\Phi({\bm p},1,0)
&=&
{1 \over 2}
\Bigl[
\langle 
c_{{\bm p},\uparrow} c_{-{\bm p},\downarrow}
\rangle
+
\langle
c_{{\bm p},\downarrow} c_{-{\bm p},\uparrow}
\rangle
\Bigr]
=-
{1 \over 2}
\Bigl[
{\hat G}^{\rm MF}_{14}({\bm p},\tau=0)
+
{\hat G}^{\rm MF}_{23}({\bm p},\tau=0)
\Bigr]
\nonumber\\
&{ }&~~~~~~~~~~~~~~~~~~~~~~~~~~~~~~~~~~~~~~
=-{\lambda_z \over 2}
{p_z \over |{\bm p}_\lambda|}
\sum_{\alpha=\pm}{\alpha\Delta_{s} \over 2E_{\bm p}^\alpha}, 
\nonumber
\\
\Phi({\bm p},1,-1)&=&
\langle 
c_{{\bm p},\downarrow} c_{-{\bm p},\downarrow}
\rangle
=-{\hat G}^{\rm MF}_{24}({\bm p},\tau=0)
=-{\lambda_\perp \over 2}
{p_x+ip_y \over |{\bm p}_\lambda|}
\sum_{\alpha=\pm}{\alpha\Delta_{s} \over 2E_{\bm p}^\alpha}.
\label{eq.17}
\end{eqnarray}
Equation (\ref{eq.7}) clearly shows that the antisymmetric spin-orbit interaction induces the $p$-wave pair amplitudes $\Phi({\bm p},S=1,S_z=\pm 1,0)$. However, we emphasize that the $p$-wave superfluid 
order parameter is still absent due to the vanishing $p$-wave pairing interaction. The system is thus in the $s$-wave superfluid state which is characterized by the $s$-wave order parameter, 
\begin{equation}
\Delta_{s}=-U_{s}\sum_{\bm p}\Phi({\bm p},0,0).
\label{eq.19}
\end{equation}
\par
At $t=0$, we suddenly replace the $s$-wave pairing interaction by the $p$-wave one\cite{OhashiP,Ho,Inotani},
\begin{equation}
H_{p-\rm{wave}} = -U_{p}
\sum_{{\bm p},{\bm p}',{\bm q}}
{\bm p}\cdot{\bm p}'
c_{{\bm p}+{\bm q}/2,\sigma}^\dagger
c_{-{\bm p}+{\bm q}/2,\sigma'}^\dagger
c_{-{\bm p}'+{\bm q}/2,\sigma'}
c_{{\bm p}'+{\bm q}/2,\sigma}.
\label{eq.20}
\end{equation}
Then, while the $s$-wave order parameter immediately vanishes due to the vanishing $s$-wave interaction, a $p$-wave order parameter $\Delta_{p}({\bm p},S_z,t>0)$ become finite, so that, by definition, 
the system is in the $p$-wave superfluid state. In particular, the $p$-wave order parameter just after this manipulation ($t=+0$) is simply given by the momentum summation of the product of the 
$p$-wave pairing interaction $U_{p}$ and a $p$-wave pair amplitude $\Phi({\bm p},S=1,S_z)|_{t=-\delta}$ that has already been prepared in the $s$-wave superfluid state. For example, 
when $\lambda_\perp=0$ and $\lambda_z\ne 0$, one has
\begin{equation}
\Delta_{p}({\bm p},S_z=0,t=+0)=-U_{p}p_z\sum_{{\bm p}'}p_z'\Phi({\bm p}',1,0)|_{t=-\delta}\propto p_z.
\label{eq.21}
\end{equation}
\par
We note that, although all the pair amplitudes $\Phi({\bm p},S,S_z)$ in Eq. (\ref{eq.17}) are proportional to the $s$-wave order parameter $\Delta_{s}$, it does not mean that they immediately disappear, 
when $\Delta_{s}$ vanishes. As well known in the proximity effect\cite{deGennes,Ohashi1996}, even when the pair amplitude penetrates into the normal metal with no interaction, it continues to exist until 
they are destroyed by external perturbations, such as thermal fluctuations and impurity scatterings. Thus, in the present case, the pair amplitudes in Eq. (\ref{eq.17}) should be regarded as 
the ``initial values'', in considering the time evolutions of $\Phi({\bm p},S,S_z,t\ge 0)$ after the sudden change of the pairing interaction from the $s$-wave one to the $p$-wave one.
\par
To evaluate the strength of the $p$-wave pair amplitude, the condensate fraction\cite{Yang,Leggett,Salanich,Giorgini,Fukushima} is a useful quantity. It is deeply related to the pair amplitude and 
physically describes the number of Bose-condensed Cooper pairs. In addition, it has widely been used in detecting the superfluid phase transition in ultracold Fermi gases\cite{Regal,Zwierlein1,Kinast,Bartenstein}. 
In the present spin-orbit coupled case, the total condensate fraction $N_{\rm c}^{\rm t}$ is given by\cite{Leggett2}
\begin{equation}
N^{\rm t}_{\rm c}={1 \over 2}
\sum_{{\bm p},\sigma,\sigma'}
|\langle c_{{\bm p},\sigma}c_{-{\bm p},\sigma'}\rangle|^2
=N_{\rm c}(S=0,S_z=0)+\sum_{S_z=-1}^1N_{\rm c}(S=1,S_z).
\label{eq.22}
\end{equation}
Here,
\begin{equation}
N_{\rm c}(S=0,S_z=0)=\sum_{\bm p}|\Phi({\bm p},0,0)|^2=
{\Delta_{s}^2 \over 16}
\sum_{\bm p}
\Bigl(
\sum_{\alpha=\pm}{1 \over E_{\bm p}^\alpha}
\Bigr)^2
\label{eq.23}
\end{equation}
is the $s$-wave condensate fraction. Equation (\ref{eq.22}) also involves the $p$-wave components $N_{\rm c}(S=1,S_z=\pm 1, 0)$, given by
\begin{eqnarray}
N_{\rm c}(1,1)&=&\frac{1}{2}\sum_{\bm p}|\Phi({\bm p},1,1)|^2
={\Delta_{s}^2 \over 32}
\sum_{\bm p}
{\lambda_\perp^2(p_x^2+p_y^2) \over |{\bm p}_\lambda|^2}
\left(
\sum_{\alpha=\pm}
{\alpha \over E_{\bm p}^\alpha}
\right)^2,
\nonumber
\\
N_{\rm c}(1,0)&=&\sum_{\bm p}|\Phi({\bm p},1,0)|^2
={\Delta_{s}^2 \over 16}
\sum_{\bm p}
{\lambda_z^2p_z^2 \over |{\bm p}_\lambda|^2}
\left(
\sum_{\alpha=\pm}
{\alpha \over E_{\bm p}^\alpha}
\right)^2,
\nonumber
\\
N_{\rm c}(1,-1)&=&\frac{1}{2}\sum_{\bm p}|\Phi({\bm p},1,-1)|^2
={\Delta_{s}^2 \over 32}
\sum_{\bm p}
{\lambda_\perp^2(p_x^2+p_y^2) \over |{\bm p}_\lambda|^2}
\left(
\sum_{\alpha=\pm}
{\alpha \over E_{\bm p}^\alpha}
\right)^2.
\label{eq.24}
\end{eqnarray}
Substituting Eqs. (\ref{eq.23}) and (\ref{eq.24}) into Eq. (\ref{eq.22}), one has
\begin{equation}
N_{\rm c}^{\rm t}={\Delta_{\rm s}^2 \over 8}\sum_{{\bm p},\alpha=\pm}
\left(
{1 \over E_{\bm p}^\alpha} 
\right)^2.
\label{eq.25}
\end{equation}
In Sec. III, we will numerically calculate $N_{\rm c}(S=1,S_z=\pm1,0)$ to examine how large the $p$-wave components are induced by a spin-orbit interaction.
\par
\par
\subsection{Superfluid phase transition temperature and effects of spin-orbit interaction}
\par
To evaluate $T_{\rm c}$ and effects of a spin-orbit interaction in the BCS-BEC crossover region, we include pairing fluctuations within the framework of the strong-coupling theory developed 
by Nozi\`eres and Schmitt-Rink\cite{NSR}. In the functional integral formalism, the $T_{\rm c}$ equation in the NSR scheme is immediately obtained from the saddle point condition, 
$\delta S_{\rm eff}/\delta\Delta^*(x)|_{\Delta(x)=\Delta^*(x)=0}=0$\cite{SadeMelo}. After the renormalization, it is given by, 
\begin{equation}
1=-{4\pi a_{s} \over m}
\sum_{\bm p}
\Bigl[{1 \over 2}
\sum_{\alpha=\pm}
{1 \over 2\xi_{\bm p}^\alpha}
\tanh{\xi_{\bm p}^{\alpha} \over 2T}
-
{1 \over 2\varepsilon_{\bm p}}
\Bigr].
\label{eq.26}
\end{equation}
As usual, we solve the $T_{\rm c}$ equation (\ref{eq.26}), together with the equation for the number $N$ of Fermi atoms, to self-consistently determine $T_{\rm c}$ and the Fermi chemical potential $\mu$. 
The NSR number equation includes pairing fluctuations within the Gaussian fluctuation level, which is derived from the identity $N=-\partial\Omega_{\rm NSR}/\partial\mu$. The NSR thermodynamic potential 
$\Omega_{\rm NSR}$ is obtained by expanding the action $S_{\rm eff}$ in Eq. (\ref{eq.9}) around $\Delta(x)=0$ to the quadratic order, which is followed by carrying out functional integrals with respect to 
$\Delta(x)$ and $\Delta^*(x)$. The result is
\begin{equation}
N=N_{\rm free}-
T{\partial \over \partial \mu}
\sum_{{\bm q},\nu_n}\ln
\Bigl[
1+{4\pi a_{s} \over m}
\Bigl[
\Pi({\bm q},i\nu_n)-\sum_{\bm p}{1 \over 2\varepsilon_{\bm p}}
\Bigr]
\Bigr]e^{i\nu_{n}\delta},
\label{eq.27}
\end{equation}
where we have eliminated the ultraviolet divergence by employing the renormalization prescription\cite{SadeMelo} and $\delta$ is infinitely small positive real number. In Eq. (\ref{eq.27}), 
$N_{\rm free}=\sum_{{\bm p},\alpha=\pm}f(\xi_{\bm p}^\alpha)$ is the number of Fermi atoms in the absence of the pairing interaction $U_{s}$, where $f(x)$ is the Fermi distribution function. 
The second term in Eq. (\ref{eq.27}) describes effects of pairing fluctuations, where 
\begin{equation}
\Pi({\bm q},i\nu_n)
={1 \over 4}\sum_{{\bm p},\alpha,\alpha'=\pm}
{1-f(\xi_{{\bm p}+{\bm q}/2}^\alpha)-f(\xi_{{\bm p}-{\bm q}/2}^{\alpha'})
\over 
\xi_{{\bm p}+{\bm q}/2}^\alpha+\xi_{{\bm p}-{\bm q}/2}^{\alpha'}-i\nu_n
}
\Bigl[
1+\alpha\alpha'
{
({\bm p}_\lambda+{\bm q}_\lambda/2)\cdot({\bm p}_\lambda-{\bm q}_\lambda/2)
\over
|{\bm p}_\lambda+{\bm q}_\lambda/2||{\bm p}_\lambda-{\bm q}_\lambda/2|
}
\Bigr]
\label{eq.29}
\end{equation}
is the lowest-order pair-correlation function in terms of the pairing interaction (where $\nu_n$ is the boson Matsubara frequency).
\par
\begin{figure*}[t]
\begin{center}
\includegraphics[width=0.5\linewidth,keepaspectratio]{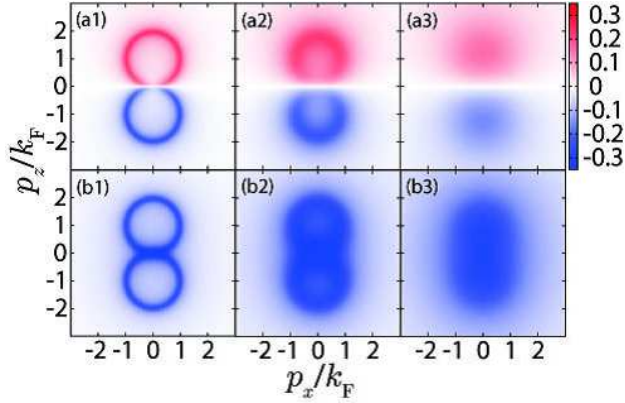}
\caption{(Color online) Calculated intensity of pair amplitude $\Phi({\bm p},S,S_z)$ at $T=0$. We set $\lambda_\perp=0$ and $\lambda_z/v_{\rm F}=1$ (where $v_{\rm F}=k_{\rm F}/m$ is the Fermi velocity 
with $k_{\rm F}$ being the Fermi momentum of a free Fermi gas in the absence of spin-orbit interaction). The upper and lower panels show the $p$-wave component $\Phi({\bm p},S=1,S_z=0)$ and 
the $s$-wave component $\Phi({\bm p},S=0,S_z=0)$, respectively. (a1) and (b1) $(k_{\rm F}a_{s})^{-1}=-1$. (a2) and (b2) $(k_{\rm F}a_{s})^{-1}=0$. (a3) and (b3) $(k_{\rm F}a_{s})^{-1}=1$. 
In these panels, we take $p_y=0$. }
\label{fig1}
\end{center}
\end{figure*}

\par
\section{$p$-wave pair amplitude induced by antisymmetric spin-orbit interaction}
\par
\subsection{single-component spin-orbit interaction ($\lambda_\perp=0,~\lambda_z\ne 0$)}
\par
Figure \ref{fig1} shows the pair amplitude $\Phi({\bm p},S,S_z)$ in a spin-orbit coupled $s$-wave superfluid Fermi gas at $T=0$ in the case of $\lambda_\perp=0$ and $\lambda_z\ne 0$. Besides the $s$-wave pair amplitude 
shown in Figs. \ref{fig1}(b1)-(b3), Figs. \ref{fig1}(a1)-(a3) show that this single-component spin-orbit interaction induces the $p$-wave component with $(S,S_z)=(1,0)$. While the $s$-wave component does not change 
its sign (See panels (b1)-(b3).), panels (a1)-(a3) clearly show the $p_z$-wave symmetry, as expected from the expression for $\Phi({\bm p},S=1,S_z=0)$ in Eq. (\ref{eq.17}).
\par
\begin{figure*}[t]
\begin{center}
\includegraphics[width=0.3\linewidth,keepaspectratio]{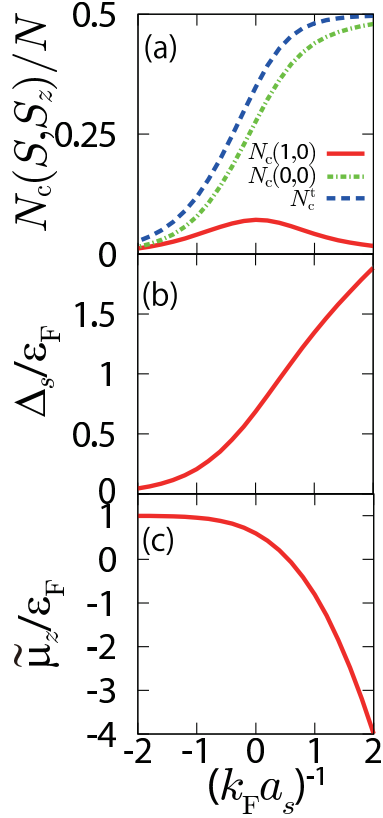}
\caption{(Color online) (a) Calculated condensate fraction $N_{\rm c}(S,S_z)$ in a spin-orbit coupled $s$-wave superfluid Fermi gas at $T=0$. We take $\lambda_z/v_{\rm F}=1$ and $\lambda_\perp=0$. In this 
single component case, one finds that $N_{\rm c}(1,\pm 1)=0$. $N_{\rm c}^{\rm t}$ is the total condensate fraction in  Eq. (\ref{eq.22}). (b) $s$-wave superfluid order parameter $\Delta_{s}$, normalized by 
the Fermi energy $\varepsilon_{\rm F}=k_{\rm F}^2/(2m)$. (c) Effective Fermi chemical potential ${\tilde \mu}_z=\mu+m\lambda_z^2/2$. 
}
\label{fig2}
\end{center}
\end{figure*}

In the weak-coupling BCS regime, one sees in Figs. \ref{fig1}(a1) and (b1) that both the $p$-wave and $s$-wave pair amplitudes are large around two circles. This is simply because, in the absence of the $s$-wave pairing 
interaction $U_{s}$, the present spin-orbit interaction gives two single-particle dispersions
\begin{equation}
\xi_{\bm p}^{\pm}={p_\perp^2 \over 2m}+{(p_z\pm m\lambda_z)^2 \over 2m}
-{\tilde \mu}_z,
\label{eq.app1}
\end{equation}
where $p_\perp^2=p_x^2+p_y^2$ and ${\tilde \mu}_z=\mu+m\lambda_z^2/2$ is an effective Fermi chemical potential. These bands give two Fermi surfaces that are centered at ${\bm p}=(0,0,\pm m\lambda_z)$ with the radius 
being equal to the Fermi momentum $k_{\rm F}$ in the absence of the spin-orbit interaction. In the weak-coupling BCS regime, since atoms near these two Fermi surfaces dominantly contribute to the Cooper-pair formation, 
the pair amplitude becomes large around them, as seen in Figs. \ref{fig1}(a1) and (b1). Since atoms away from the Fermi surfaces also contribute to the pair formation when the pairing interaction is strong, 
the circular structure is obscure in Figs. \ref{fig1}(a2) and (b2). In the strong-coupling BEC regime, the effective Fermi chemical potential ${\tilde \mu}_{z}$ is negative (See Fig. \ref{fig2}(c).), so that 
the Fermi surface no longer exists. As a result, the circular structure disappears in Figs. \ref{fig1}(a3) and (b3).
\par
\par
To evaluate the $p$-wave pair amplitude in a quantitative manner, we conveniently consider the condensate fraction $N_{\rm c}(S,S_z)$ at $T=0$. In the BCS side ($(k_{\rm F}a_{s})^{-1}\lesssim 0$), 
Fig. \ref{fig2}(a) shows that the $p$-wave component $N_{\rm c}(S=1,S_z=0)$ increases with increasing the interaction strength, reflecting the increase of the magnitude of the $s$-wave 
superfluid order parameter $\Delta_{s}$ shown in Fig. \ref{fig2}(b). However, $N_{\rm c}(S=1,S_z=0)$ decreases in the BEC side ($(k_{\rm F}a_{s})^{-1}\gesim 0$), although the $s$-wave 
condensate fraction $N_{\rm c}(S=0,S_z=0)$ continues to increase. In this regime, the coupled equations (\ref{eq.14}) and (\ref{eq.15}) give,
\begin{eqnarray}
\Delta_{s}&=&\sqrt{16 \over 3\pi (k_{\rm F}a_{s})}\varepsilon_{\rm F},
\nonumber
\\
\mu &=&
-{1 \over 2ma_{s}^2}-{1 \over 2}m\lambda_z^2,
\label{eq.30}
\end{eqnarray}
so that one obtains
\begin{eqnarray}
N_{\rm c}(1,0)&=&{N \over 12}
\left({\lambda_z \over v_{\rm F}}\right)^2(k_{\rm F}a_{s})^2, 
\nonumber
\\
N_{\rm c}(0,0)&=&{N \over 2}-N_{\rm c}(1,0).
\label{eq.31}
\end{eqnarray}
This means that the strong-coupling BEC limit ($(k_{\rm F}a_{s})^{-1}\to \infty$) may be simply viewed as a gas of $N/2$ $s$-wave molecules, even in the presence of spin-orbit interaction.
\par

\begin{figure}
\begin{center}
\includegraphics[width=0.4\linewidth,keepaspectratio]{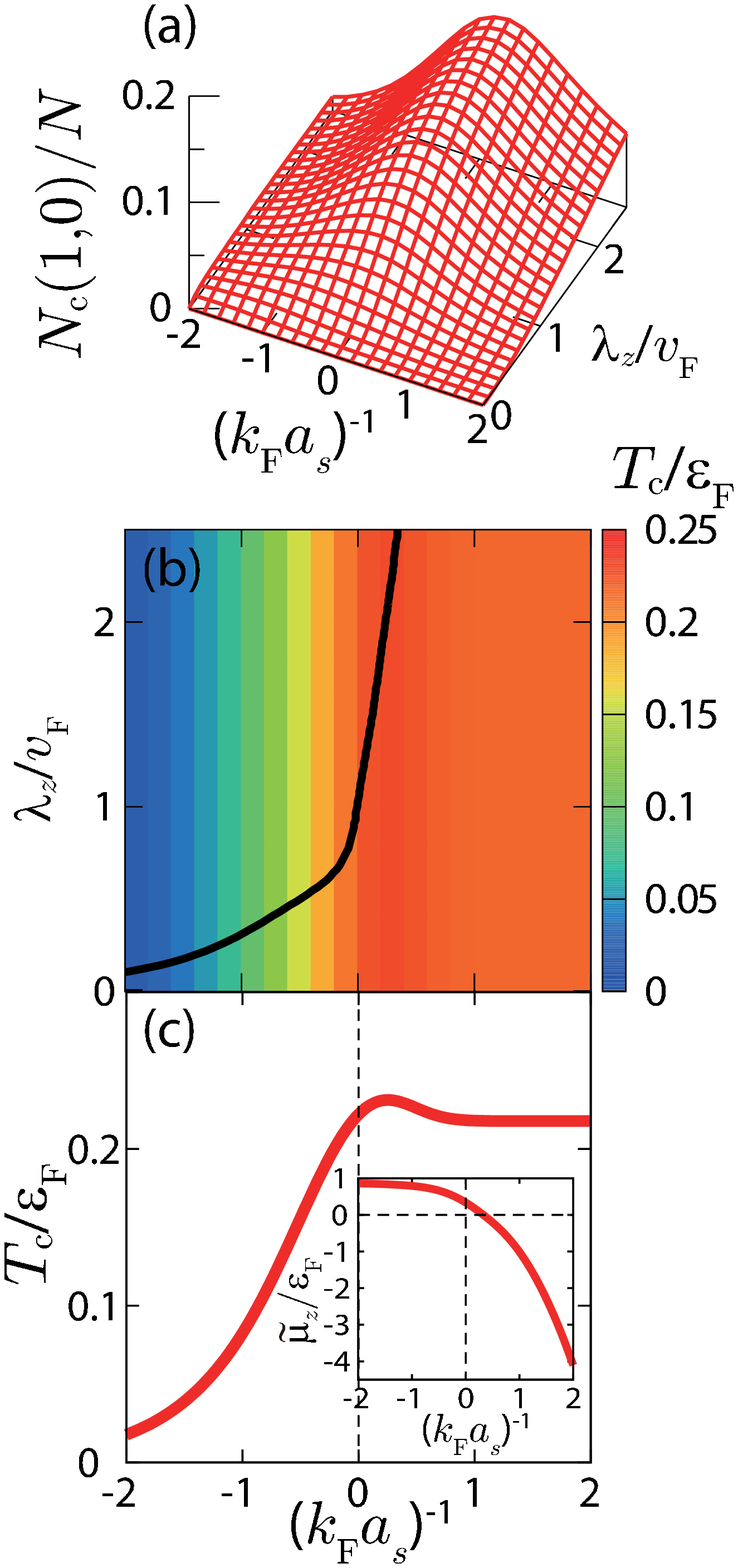}
\caption{(Color online) (a) Calculated $p$-wave condensate fraction $N_{\rm c}(S=1,S_z=0)$ at $T=0$ in the case of single-component spin-orbit interaction ($\lambda_\perp=0$). (b) Peak position of $N_{\rm c}(S=1,S_z=0)$ 
evaluated from the result shown in panel (a). The background intensity shows $T_{\rm c}$. (c) Calculated $T_{\rm c}$. The inset shows the effective Fermi chemical potential ${\tilde \mu}_z=\mu+m\lambda^2_z/2$.
}
\label{fig3}
\end{center}
\end{figure}
\par
When the spin-orbit interaction becomes strong, the magnitude of $p$-wave component $N_{\rm c}(S=1,S_z=0)$ increases, as shown in Fig. \ref{fig3}(a). However, in the present case of single-component spin-orbit interaction, 
we should note that the total condensate fraction $N_{\rm c}^{\rm t}$, as well as the $s$-wave superfluid order parameter $\Delta_{s}$, do not depend on the spin-orbit coupling strength $\lambda_z$. Indeed, when 
$\lambda_\perp=0$ and $\lambda_z\ne 0$, the coupled equations (\ref{eq.14}) and (\ref{eq.15}) are reduced to the ordinary BCS-Leggett coupled equations for a $s$-wave Fermi superfluid {\it with no spin-orbit interaction}, 
as 
\begin{eqnarray}
1&=&-{4\pi a_{s} \over m}
\sum_{\bm p}
\left[
{1 \over 2\sqrt{{\tilde \xi}_{\bm p}^2+\Delta_{s}^2}}
-{1 \over 2\varepsilon_{\bm p}}
\right],
\nonumber
\\
N&=&\sum_{\bm p}
\left[
1-
{{\tilde \xi}_{\bm p}
\over 
\sqrt{{\tilde \xi}_{\bm p}^2+\Delta_{s}^2}
}
\right],
\label{eq.32}
\end{eqnarray}
where ${\tilde \xi}_{\bm p}=\varepsilon_{\bm p}-{\tilde \mu}_z$. Thus, the self-consistent solutions $(\Delta_{s},{\tilde \mu}_{z})$ shown in Figs. \ref{fig2}(b) and (c) are independent of $\lambda_z$. As a result, 
the total condensate fraction $N_{\rm c}^{\rm t}$ in Eq. (\ref{eq.25}) is also $\lambda_z$-independent as
\begin{equation}
{N_{\rm c}^{\rm t} \over N}=
{\Delta_{s}^2 \over 4N}
\sum_{\bm p}
{1 \over {\tilde \xi}_{\bm p}^2+\Delta_{s}^2}
=
{3\pi \Delta_{s} \over 16\sqrt{2}\varepsilon_{\rm F}^{3/2}}
\sqrt{{\tilde \mu}_{z}+\sqrt{{\tilde \mu}_z^2+\Delta_{s}^2}},
\label{eq.33}
\end{equation}
where $\varepsilon_{\rm F}$ is the Fermi energy for a free Fermi gas without spin-orbit interaction. As a result, for a given interaction strength $(k_{\rm F}a_{s})^{-1}$, the $s$-wave condensate fraction 
$N_{\rm c}(S=0,S_z=0)=N_{\rm c}^{\rm t}-N_{\rm c}(S=1,S_z=0)$ becomes small when the spin-orbit interaction becomes strong, in spite of the fact that the $s$-wave superfluid order parameter $\Delta_{s}$ remains unchanged.
\par

\par
Evaluating the peak position of the $p$-wave condensate fraction $N_{\rm c}(S=1,S_z=0)$ from Fig. \ref{fig3}(a), we obtain Fig. \ref{fig3}(b). Recently, the single-component spin-orbit interaction with 
$0.5\lesssim \lambda_z/v_{\rm F}\lesssim 1$ has been realized in a $^{40}$K Fermi gas\cite{K-SOC}. Keeping this in mind, we find from Fig. \ref{fig3}(b) that the region
\begin{equation}
\label{eq.34}
(k_{\rm F}a_{s})^{-1}\simeq 0,~~\lambda_z/v_{\rm F}\simeq 1,
\end{equation}
 is suitable for the preparation of large $p$-wave pair amplitude. As an example, at $(k_{\textrm{F}}a_{s})^{-1}=0$ and $\lambda/v_{\rm{F}}=1$, one obtains 
\begin{equation}
{N_{\rm c}(S=1,S_z=0) \over N}\simeq 0.07.
\label{eq.35}
\end{equation}
In this case, just after the sudden change of the pairing interaction from the $s$-wave one to the $p$-wave one in Eq. (\ref{eq.20}), we expect the $p$-wave superfluid state with the $p$-wave condensate fraction 
being equal to Eq. (\ref{eq.35}). This $p$-wave superfluid state with $\Delta_p({\bm p},S_z=0)\propto p_z$ is just the so-called polar phase discussed in superfluid $^3$He\cite{Vollhardt,Mineev}. 
As usual\cite{Vollhardt,Mineev,Sigrist}, one can summarize this $p$-wave superfluid order parameter as
\begin{eqnarray}
{\hat \Delta}({\bm p})=
\left(
\begin{array}{cc}
\Delta_p({\bm p},S_z=1) &
\Delta_p({\bm p},S_z=0) \\
\Delta_p({\bm p},S_z=0) &
\Delta_p({\bm p},S_z=-1) \\
\end{array}
\right)
\sim
\left(
\begin{array}{cc}
0 &
p_z \\
p_z &
0 \\
\end{array}
\right),
\label{eq.A}
\end{eqnarray}
where we have ignored the unimportant factor that is nothing to do with the pairing symmetry in the last expression. Since the polar state has not been realized in liquid $^3$He, a single-component spin-orbit coupled 
$s$-wave superfluid Fermi gas would be useful for the realization of this unconventional pairing state.
\par
To actually prepare the $p$-wave pair amplitude in the parameter region in Eq. (\ref{eq.34}), the $s$-wave superfluid phase in this regime must be experimentally accessible. In this regard, we note that $T_{\rm c}$ 
is also $\lambda_z$-independent in the present case. Indeed, when $\lambda_\perp=0$ and $\lambda_z\ne 0$, we can completely eliminate the $\lambda_z$-dependence from the $T_{\rm c}$-equation (\ref{eq.26}) as
\begin{equation}
1=-{4\pi a_{s} \over m}
\sum_{\bm p}
\Bigl[{1 \over 2{\tilde \xi}_{\bm p}}
\tanh{{\tilde \xi}_{\bm p} \over 2T}
-
{1 \over 2\varepsilon_{\bm p}}
\Bigr],
\label{eq.36}
\end{equation}
where ${\tilde \xi}_{\bm p}$ is given below Eq. (\ref{eq.32}). In the same manner, the non-interacting part $N_{\rm free}$, as well as the pair correlation function $\Pi({\bm q},i\nu_n)$, in the number equation 
(\ref{eq.27}) can be also written in the $\lambda_z$-independent forms as, respectively,
\begin{equation}
N_{\rm free}=2\sum_{\bm p}f({\tilde \xi}_{\bm p}),
\label{eq.37}
\end{equation}
\begin{equation}
\Pi({\bm q},i\nu_n)
=\sum_{\bm p}
{1-f({\tilde \xi}_{{\bm p}+{\bm q}/2})-f({\tilde \xi}_{{\bm p}-{\bm q}/2})
\over 
{\tilde \xi}_{{\bm p}+{\bm q}/2}+{\tilde \xi}_{{\bm p}-{\bm q}/2}-i\nu_n
}.
\label{eq.38}
\end{equation}
Thus, the self-consistent solutions of the coupled equations (\ref{eq.26}) and (\ref{eq.27}) in Fig. \ref{fig3}(c) is valid for arbitrary values of the spin-orbit coupling strength $\lambda_z$. Since current experiments 
can reach the temperature region far below $T_{\rm c}$ of a unitary Fermi gas in the absence of a spin-orbit interaction\cite{Regal,Zwierlein1,Kinast,Bartenstein}, the region in Eq. (\ref{eq.34}) is also accessible 
within the current experimental technique. In addition, since the superfluid order rapidly grows in the superfluid phase below $T_{\rm c}$, when the temperature can be lowered to some extent below $T_{\rm c}$, we would 
be able to obtain the $p$-wave condensate fraction, the value of which is close to that at $T=0$ obtained in this paper.
\par
\begin{figure*}[t]
\begin{center}
\includegraphics[width=0.6\linewidth,keepaspectratio]{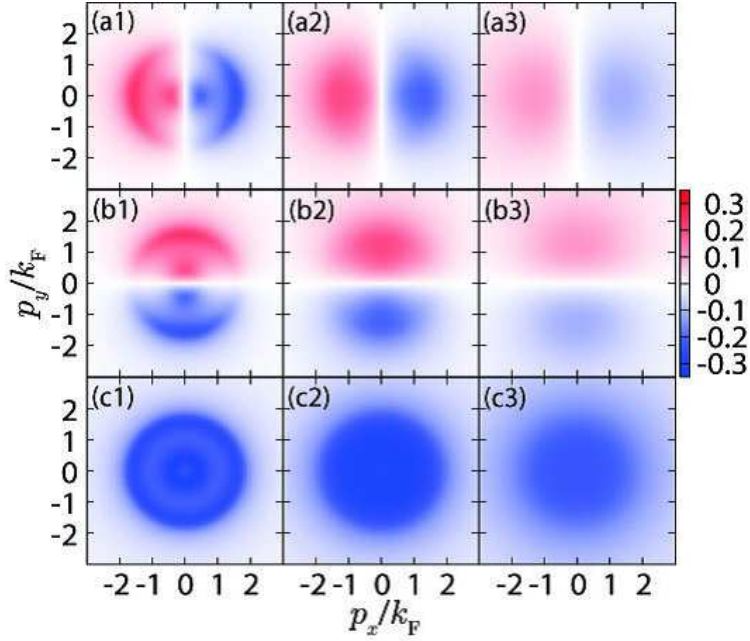}
\caption{(Color online) Intensity of the pair amplitude $\Phi({\bm p},S,S_z)$ at $T=0$ in the case of $\lambda_\perp/v_{\rm F}=1$ and $\lambda_z=0$. The upper and middle panels show ${\rm Re}[\Phi({\bm p},S=1,S_z=1)]$ 
and ${\rm Im}[\Phi({\bm p},S=1,S_z=1)]$, respectively. The lower panels show the $s$-wave pair amplitude $\Phi({\bm p},S=0,S_z=0)$. (a1)-(c1) $(k_{\rm F}a_{s})^{-1}=-1$. (a2)-(c2) $(k_{\rm F}a_{s})^{-1}=0$. 
(a3)-(c3) $(k_{\rm F}a_{s})^{-1}=1$. In these panels, we take $p_z=0$. In the present case, the $p$-wave component with $S_z=-1$ is also induced (although we do not show this here), which is simply related to 
the $S_z=+1$ component as $\Phi({\bm p},S=1,S_z=-1)=-\Phi^*({\bm p},S=1,S_z=1)$.
}
\label{fig4}
\end{center}
\end{figure*}
\par
\subsection{two-component spin-orbit interaction ($\lambda_\perp\ne 0,~\lambda_z=0$)}
\par
When $\lambda_\perp\ne 0$ and $\lambda_z=0$, the spin-orbit interaction in Eq. (\ref{eq.7}) consists of the $\sigma_x$ and $\sigma_y$ components. In this two-component case, Eq. (\ref{eq.17}) indicates that 
the $p$-wave pair amplitudes with $S_z=\pm 1$ are induced. We show the detailed momentum dependence of the $S_z=1$ component in the upper and middle panels in Fig. \ref{fig4}. Although we do not show the 
$S_z=-1$ component, it is simply related to the $S_z=1$ component as $\Phi({\bm p},S=1,S_z=-1)=-\Phi^*({\bm p},S=1,S_z=1)$. Thus, just after the $s$-wave interaction is suddenly replaced by the $p$-wave one 
in Eq. (\ref{eq.20}), one has the $p$-wave superfluid order parameters $\Delta_p({\bm p},S_z=\pm 1)$. Because $|\Delta_p({\bm p},S_z=1)|=|\Delta_p({\bm p},S_z=-1)|$, this $p$-wave superfluid phase is just 
the planar state\cite{Vollhardt,Sigrist,Mineev}. Under the matrix representation in Eq. (\ref{eq.A}), one has
\begin{eqnarray}
{\hat \Delta}({\bm p})\sim
\left(
\begin{array}{cc}
-p_x+ip_y &
0 \\
0 &
p_x+ip_y \\
\end{array}
\right).
\label{eq.A2}
\end{eqnarray}
We briefly note that the planar state has not been realized in liquid $^3$He.
\par

\begin{figure*}[t]
\begin{center}
\includegraphics[width=0.7\linewidth,keepaspectratio]{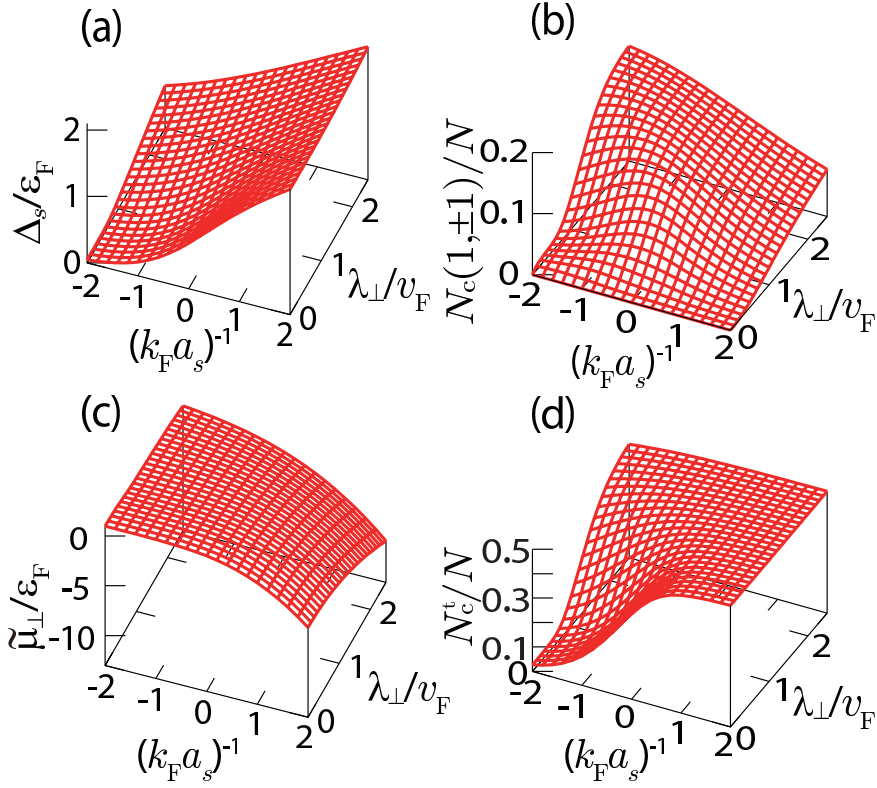}
\caption{(Color online) (a) Calculated $s$-wave superfluid order parameter $\Delta_{s}$ and (c) effective chemical potential ${\tilde \mu}_\perp=\mu+m\lambda_\perp^2/2$ at $T=0$ in the case of two-component 
spin-orbit interaction ($\lambda_\perp\ne 0, \lambda_z=0$). (b) $p$-wave condensate fraction $N_{\rm c}(S=1,S_z=\pm 1)$. (d) Total condensate fraction $N_{\rm c}^{\rm t}$.
}
\label{fig5}
\end{center}
\end{figure*}
\par
As in the single-component case, the pair amplitude dominantly appears around the Fermi surface in the weak-coupling BCS regime. When $U_{s}=0$, the present two-component spin-orbit interaction gives 
\begin{equation}
\xi_{\bm p}^\pm={(p_\perp\pm m\lambda_\perp)^2 \over 2m}+
{p_z^2 \over 2m}-{\tilde \mu}_\perp,
\label{eq.39}
\end{equation}
where ${\tilde \mu}_\perp=\mu+m\lambda_\perp^2/2$. At $p_z=0$, one obtains two Fermi surfaces, both of that are centered at ${\bm p}=0$, with the radii, 
\begin{eqnarray}
k_{\rm F}^\pm=
\left\{
\begin{array}{ll}
\sqrt{2m{\tilde \mu}_\perp}\pm m\lambda_\perp,&
{\tilde \mu}_\perp\ge m\lambda_\perp^2/2,\\
\pm\sqrt{2m{\tilde \mu}_\perp}+ m\lambda_\perp,&
{\tilde \mu}_\perp< m\lambda_\perp^2/2.\\
\end{array}
\right.
\label{band1}
\end{eqnarray}
Figures \ref{fig4}(a1)-(c1) show that both the $s$-wave and $p$-wave pair amplitudes are large around these Fermi surfaces, as expected. 
\par
In the two-component case, one cannot eliminate the $\lambda_\perp$-dependence from the BCS-Leggett coupled equations (\ref{eq.14}) and (\ref{eq.15}). As a result, the $s$-wave superfluid order parameter 
$\Delta_{s}$, as well as the effective Fermi chemical potential ${\tilde \mu}_\perp$, depend on the spin-orbit coupling strength $\lambda_\perp$, as shown in Figs. \ref{fig5}(a) and (c). Because of this, 
we see in Fig. \ref{fig5}(b) and (d) that, not only the $p$-wave condensate fraction $N_{\rm c}(S=1,S_z=\pm 1)$, but also the total condensate fraction $N_{\rm c}^{\rm t}$ is affected by the spin-orbit 
interaction, which is different from the single-component case discussed in the previous subsection. The $\lambda_\perp$ dependence of $N_{\rm c}^{\rm t}$ can also be confirmed by analytically carrying out 
the momentum summation in Eq. (\ref{eq.25}) as
\begin{equation}
{N_{\rm c}^{\rm t} \over N}
=
{3\pi\Delta_{s} \over 16\sqrt{2}\varepsilon_{\rm F}^{3/2}}
\left[
\sqrt{\mu+\sqrt{\mu^{2}+\Delta_{s}^2}}
+
{\sqrt{m}\lambda_\perp \over 2}
\arccos
{
\sqrt{\mu^2+\Delta_{s}^2}-m\lambda_\perp^2/2
\over
\sqrt{{\tilde \mu}_\perp^2+\Delta_{s}^2}
}
\right].
\label{eq.40}
\end{equation}
\par
\begin{figure*}[t]
\begin{center}
\includegraphics[width=0.4\linewidth,keepaspectratio]{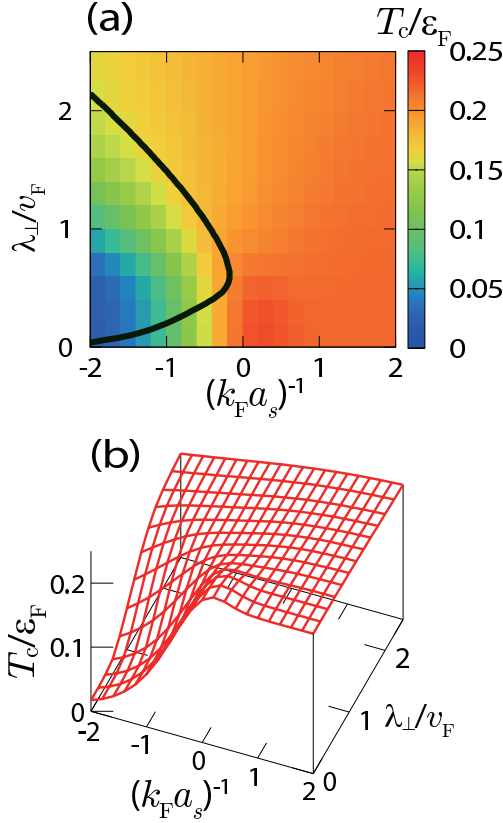}
\caption{(Color online) (a) Peak position of $p$-wave condensate fraction $N_{\rm c}(S=1,S_z=\pm 1)$ at $T=0$ in the case of two-component spin-orbit interaction. The background intensity represents the 
magnitude of $T_{\rm c}$. For convenience, we also show the three-dimensional plot of $T_{\rm c}$ in panel (b).
}
\label{fig6}
\end{center}
\end{figure*}

Figure \ref{fig5}(d) shows that the total condensate fraction $N_{\rm c}^{\rm t}$ in the BCS side ($(k_{\rm F}a_{s})^{-1}\lesssim 0$) is remarkably enhanced by the spin-orbit interaction, to be comparable to 
the value in the BCS regime with $\lambda_\perp=0$. Reflecting this enhancement, we find from Fig. \ref{fig5}(b) that the $p$-wave component $N_{\rm c}(S=1,S_z=\pm 1)$ is also enhanced in the BCS regime, 
when $\lambda_\perp$ is large. Thus, as shown in Fig. \ref{fig6}(a), the peak position of $N_{\rm c}(S=1,S_z=\pm 1)$ shifts to the BCS side, compared to the single component case shown in Fig. \ref{fig3}(b).
\par
\par
Although the superfluid phase transition temperature $T_{\rm c}$ is usually low in the weak-coupling BCS regime, Fig. \ref{fig6}(b) shows that $T_{\rm c}$ around the ``optimal region" (peak line in Fig. \ref{fig6}(a)) 
in the BCS regime is also enhanced by the spin-orbit interaction. Since we need to reach the $s$-wave superfluid phase in order to prepare the $p$-wave pair amplitude, this enhancement of $T_{\rm c}$ is favorable for 
our purpose. We briefly note that the enhancement of $T_{\rm c}$ by a Rashba-type spin-orbit interaction has recently been pointed out\cite{Zheng}.  
\par
\begin{figure*}[t]
\begin{center}
\includegraphics[width=0.5\linewidth,keepaspectratio]{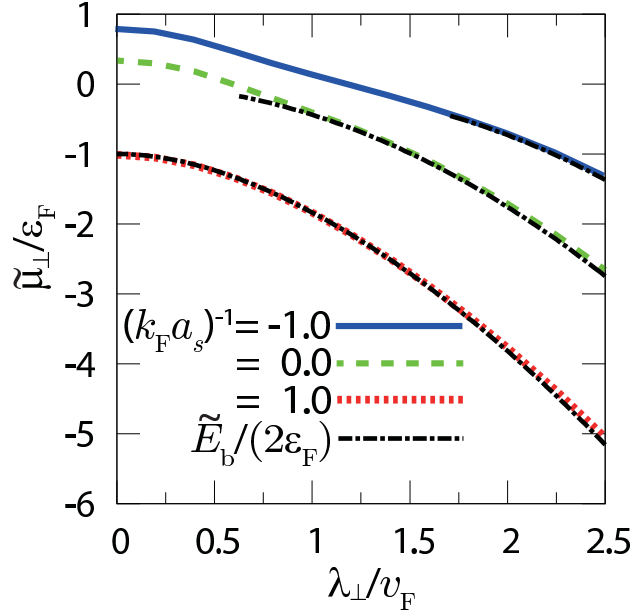}
\caption{(Color online) Calculated effective chemical potential ${\tilde \mu}_\perp=\mu+m\lambda_\perp^2/2$ at $T_{\rm c}$ as a function of the spin-orbit coupling strength $\lambda_\perp$. 
${\tilde E}_{\rm b}/2=E_{\rm b}/2+m\lambda_\perp^2/2$, where $E_{\rm b}$ is the binding energy of a two-body bound state, determined from Eq. (\ref{eq.bind}).
}
\label{fig7}
\end{center}
\end{figure*}
\par
The reason for the enhancement of the condensate fraction, as well as $T_{\rm c}$, in the weak-coupling BCS regime is the formation of two-body bound molecules\cite{2B1,Zheng,rashbon1,rashbon2,2B2,2B3}, which are also 
referred to as rashbons in the literature. In the present case, the degeneracy of the lowest energy level in the lower band $\xi_{\bm p}^-$ in Eq. (\ref{eq.39}) leads to a two-dimensional-like density of states around 
the bottom of this band. This ``low-dimensional effect" stabilizes a two-body bound state (rashbon), even in the weak-coupling BCS regime\cite{2B1,Zheng,rashbon1,rashbon2,2B2,2B3}, where such a two-body bound state 
does not appear in the ordinary three-dimensional system. Then, the superfluid phase transition in the strong spin-orbit coupling regime is dominated by the BEC of rashbons, giving a high $T_{\rm c}$ even in the BCS regime. 
This is similar to the case of the ordinary strong-coupling BEC regime of an ultracold Fermi gas\cite{Eagles,Leggett,Leggett2,NSR,SadeMelo,Ohashi,Strinati,Levin,Giorginia,Bloch,Zwerger}, where the superfluid phase 
transition is dominated by tightly bound molecules that are formed by a strong pairing interaction. Indeed, as shown in Fig. \ref{fig7}, when the spin-orbit coupling strength $\lambda_\perp$ increases, the effective 
Fermi chemical potential ${\tilde \mu}_\perp=\mu+m\lambda_\perp^2/2$ at $T_{\rm c}$ becomes negative, to approach ${\tilde E}_{\rm b}/2=E_{\rm b}/2+m\lambda_\perp^2/2$, where $E_{\rm b}$ is the binding energy of 
a two-body bound state, determined from the equation\cite{2B1,Zheng,rashbon1,rashbon2,2B2,2B3},
\par
\begin{equation}
1=-
{4\pi a_{s} \over m}
\sum_{\bm p}
\left[
{1 \over 2}
\sum_{\alpha=\pm}
{1 \over 
2(\varepsilon_{\bm p}+\alpha\lambda_\perp p_\perp)-E_{\rm b}}-{1 \over 2\varepsilon_{\bm p}}
\right]
\quad  (E_{\rm b} < -m\lambda_{\perp}^{2}/2).
\label{eq.bind}
\end{equation}
Since the chemical potential physically describes energy to add a particle to the system, Fig. \ref{fig7} indicates that most Fermi atoms form two-body bound molecules in the strong spin-orbit coupling regime.
\par
\begin{figure*}[t]
\begin{center}
\includegraphics[width=0.5\linewidth,keepaspectratio]{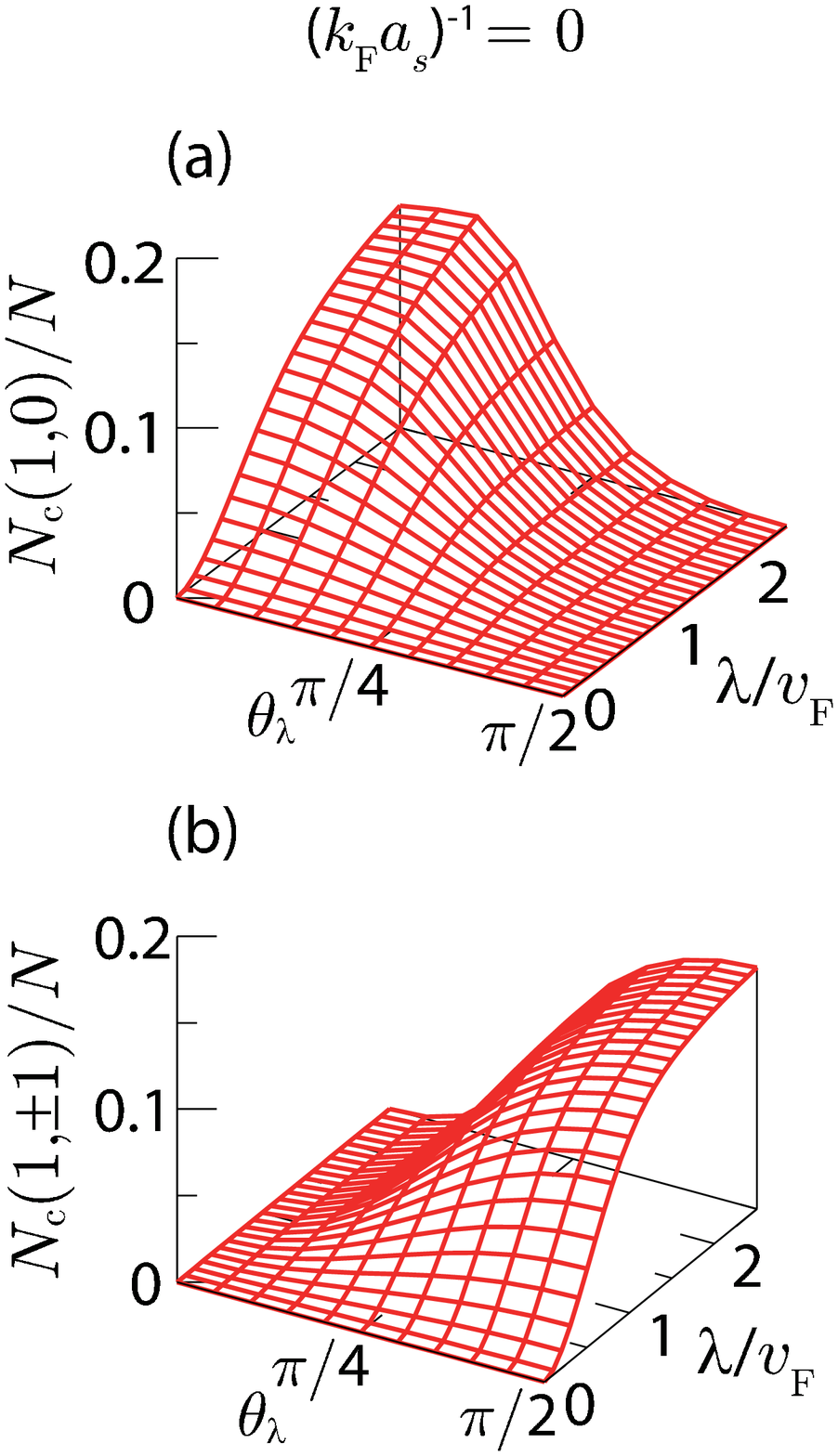}
\caption{(Color online) Calculated $p$-wave condensate fraction at $T=0$ in the case of three component spin-orbit interaction $(\lambda_\perp,\lambda_z)=\lambda(\sin\theta_\lambda,\cos\theta_\lambda)$. 
We set $(k_{\rm F}a_{s})^{-1}=0$. (a) $N_{\rm c}(S=1,S_z=0)$. (b) $N_{\rm c}(S=1,S_z=\pm 1)$. In this figure, $\theta_\lambda=0$ and $\theta_\lambda=\pi/2$, respectively, correspond to the single-component 
and two-component spin-orbit interaction, discussed in the previous subsections.
}
\label{fig8}
\end{center}
\end{figure*}
\par
\begin{figure*}[t]
\begin{center}
\includegraphics[width=0.4\linewidth,keepaspectratio]{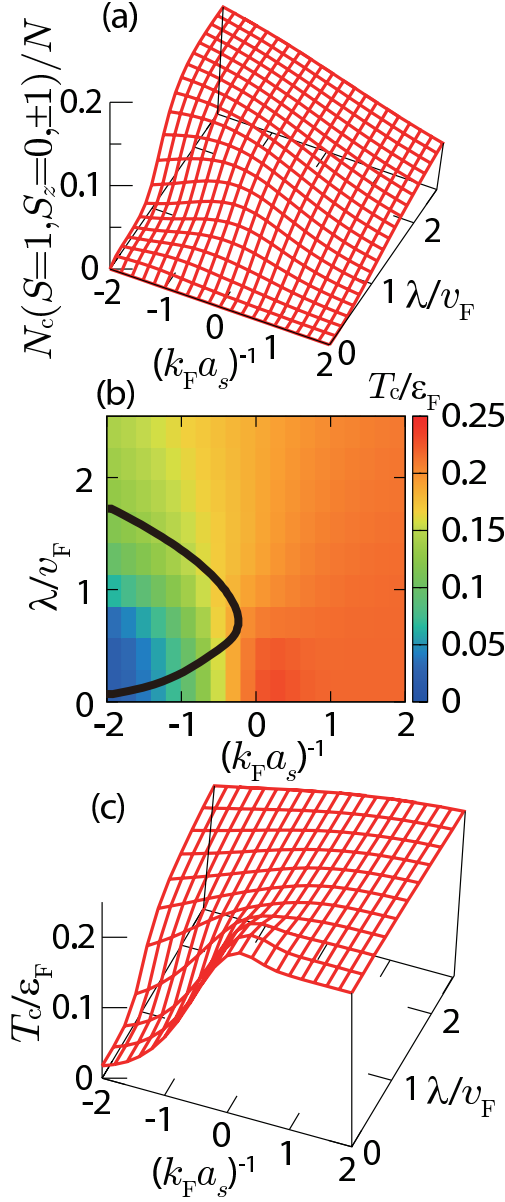}
\caption{(Color online) (a) $p$-wave condensate fraction $N_{\rm c}(S=1,S_z=0,\pm 1)$ at $T=0$ in the case of isotropic spin-orbit interaction $\lambda_\perp=\lambda_z=\lambda/\sqrt{2}$ ($\theta_\lambda=\pi/4$). 
(b) Peak position of $N_{\rm c}(S=1,S_z=0,\pm 1)$ evaluated from panel (a). The background intensity shows $T_{\rm c}$, which is also shown in panel (c) for clarity.
}
\label{fig9}
\end{center}
\end{figure*}
\par
\subsection{Three component spin-orbit interaction ($\lambda_\perp\ne 0,~\lambda_z\ne 0$)}
\par
When the spin-orbit interaction has all the $\sigma_x$, $\sigma_y$, and $\sigma_z$ components [$(\lambda_\perp,\lambda_z)=\lambda(\cos\theta_\lambda,\sin\theta_\lambda)$ ($0\le\theta_\lambda\le\pi/2)$], all the 
$p$-wave condensate fractions, $N_{\rm c}(S=1,S_z=\pm 1, 0)$, are induced, as shown in Fig. \ref{fig8}. 
\par
When the spin-orbit interaction is isotropic ($\lambda_\perp=\lambda_z$ or $\theta_\lambda=\pi/4$), Figs. \ref{fig9}(a) and (b) indicate that the BCS side is suitable for our purpose. Although we do not explicitly 
show the result here, the total condensate fraction, which is analytically given by
\begin{equation}
{N_{\rm c}^{\rm t} \over N}
=
{3\pi \Delta_{s} \over 16\sqrt{2}\varepsilon_{\rm F}^{3/2}}
\sqrt{{\tilde \mu}+\sqrt{{\tilde \mu}^2+\Delta_{s}^2}}
\left[
1+
{m\lambda^2/4
\over 
\sqrt{{\tilde \mu}^2+\Delta_{s}^2}}
\right], 
\label{eq.A4}
\end{equation}
(where ${\tilde \mu}=\mu+m\lambda^2/4$) is also enhanced in the BCS side by the rashbon formation discussed in the previous subsection. In addition, $T_{\rm c}$ is also enhanced by the same mechanism 
(See Fig. \ref{fig9}(c)), so that this regime is still considered to experimentally be accessible. In this case, when the $s$-wave pairing interaction is suddenly replaced by the $p$-wave one in Eq. (\ref{eq.20}), 
as the initial $p$-wave state, we can prepare the BW (Balian-Werthamer) superfluid order parameter\cite{Vollhardt,Mineev,Sigrist}, having the form
\begin{eqnarray}
{\hat \Delta}({\bm p})\sim
\left(
\begin{array}{cc}
-p_x+ip_y &
p_z \\
p_z &
p_x+ip_y \\
\end{array}
\right).
\label{eq.A3}
\end{eqnarray}
The BW state has been realized in superfluid $^3$He\cite{Vollhardt}.
\par
\par
\section{Summary}
\par
To summarize, we have discussed a possible idea to realize a $p$-wave superfluid Fermi gas. In contrast to the ordinary approach where one tries to cool down a Fermi gas with a strong $p$-wave pairing interaction, 
our idea consists of two stages. In the first stage, we only prepare a $p$-wave pair amplitude by, not using a $p$-wave interaction, but using the phenomenon that a $p$-wave pair amplitude is induced in a 
$s$-wave Fermi superfluid with broken inversion symmetry by an antisymmetric spin-orbit interaction. Then, in the second stage, we suddenly replace the $s$-wave interaction by a $p$-wave one, to produce the 
$p$-wave superfluid order parameter that is essentially given by the product of the $p$-wave interaction and the $p$-wave pair amplitude which has been prepared in the first stage. In this paper, we have 
assessed the first stage of our idea, by evaluating how large the $p$-wave pair amplitude can be induced in a spin-orbit coupled $s$-wave superfluid state. We clarified the region where a large $p$-wave pair 
amplitude is obtained, in the phase diagram of a Fermi gas with respect to the strengths of the $s$-wave pairing interaction and the spin-orbit coupling. Within the framework of the NSR theory, we also confirmed 
that the $s$-wave superfluid phase in this optimal region is accessible within the current experimental technique.
\par
The key of our idea is that the pairing symmetry of a Fermi superfluid is just the symmetry of the superfluid order parameter, which is essentially given by the product of a pairing interaction and a pair amplitude. 
Thus, even when a $p$-wave pair amplitude is induced in a $s$-wave superfluid Fermi gas, the system is still in the $s$-wave superfluid state because of the vanishing $p$-wave interaction. This also implies 
the possibility that one may artificially produce a $p$-wave Fermi superfluid by separately preparing a $p$-wave pair amplitude and a $p$-wave interaction. Our idea just uses this possibility, that is, we separately 
prepare these two quantity by using two sophisticated techniques developed in cold atom physics: the artificial gauge field and the tunable pairing interaction associated with a Feshbach resonance.
\par
At present, all the experiments aiming a $p$-wave superfluid Fermi gas uses a $p$-wave pairing interaction from the beginning. While this approach is straightforward, it seems suffering from the short lifetime 
of $p$-wave molecules, as well as the particle loss by the $p$-wave interaction. In contrast, since our idea does not rely on the $p$-wave interaction in the first stage, this difficulty can be avoided to some extent. 
In addition, since we can start from the situation with a finite value of the $p$-wave superfluid order parameter and a finite value of the $p$-wave condensate fraction, even when the $p$-wave interaction eventually 
destroys this superfluid state, we expect to be able to realize a $p$-wave superfluid Fermi gas for a while, after the $p$-wave interaction is introduced.
\par
In this paper, we have only discussed the first stage of our idea. Since the $p$-wave superfluid state which is artificially produced would be in the non-equilibrium state, as the next step, we need to examine 
how this $p$-wave state relaxes into the thermodynamically stable state, after the $p$-wave interaction is turned on. In addition, since there are various $p$-wave superfluid phases, such as the BW phase and polar phase, 
the initial $p$-wave superfluid state may not be the most stable state for a given strength of a $p$-wave interaction and a given value of temperature. Thus, in order to soon reach the thermodynamically stable state, 
one need to clarify the optimal condition for the initial $p$-wave pairing state. In our future papers, we will discuss these interesting problems existing in the second stage of our idea. Since both the spin and 
orbital degrees of freedom are active in a $p$-wave Fermi superfluid, the realization of a $p$-wave superfluid Fermi gas would enable us to discuss much richer physics than the case of the $s$-wave superfluid Fermi gas. 
Since current experiments toward the realization of a unconventional superfluid Fermi gas faces various difficulties, our results would be useful for the exploration of a route to accomplish this exciting challenge 
in cold Fermi gas physics.
\par
\begin{acknowledgments}
We thank D. Inotani, Y. Endo, R. Hanai, H. Tajima, M. Matsumoto and P. van Wyk for discussions and comments. YO thanks T. Mukaiyama for giving him useful information about the current experimental situation. This work was 
supported by KiPAS project in Keio University. YO was also supported by Grant-in-Aid for Scientific research from MEXT in Japan (25105511, 25400418, 15H00840).
\end{acknowledgments}
\newpage
\par

\end{document}